\documentclass[12pt]{article}
\newlength{\vshift}
\input epsf.tex
\newlength{\hshift}

\usepackage{amssymb}
\usepackage{amsmath}
\usepackage{amsthm}
\usepackage{amsopn}

\def\comment#1{}

\def\beq{\begin{equation}}
\def\eeq{\end{equation}}
\def\bea{\begin{eqnarray}}
\def\eea{\end{eqnarray}}

\usepackage{ulem}
\usepackage{cancel}
\usepackage{color}

\begin{document}

 \vspace*{3cm}

\begin{center}

{ \Large \bf{Euler-Heisenberg Lagrangian and photon circular polarization}}

\vskip 4em

{
 {\bf Iman Motie$^{\it a}$} and {\bf She-Sheng Xue$^{\it b}$} \footnote{e-mail: i.motie@ph.iut.ac.ir and xue@icra.it }
}

\vskip 1em {\it a) Department of Physics, Mashhad Branch, Islamic
Azad University, Iran\\} {\it b) ICRANet, P.zza della Repubblica
10, I-65122 Pescara, \& Physics Department, University of Rome
``La Sapienza'', Italy}
\end{center}

 \vspace*{1.5cm}

\begin{abstract}
Considering the effective Euler-Heisenberg Lagrangian, i.e.,
non-linear photon-photon interactions, we study the circular
polarization of electromagnetic radiation based
on the time-evolution of Stokes parameters. To the leading order,
we solve the Quantum Boltzmann Equation for the density matrix
describing an ensemble of photons in the space of energy-momentum
and polarization states, and calculate
the intensity of circular polarizations. Applying these results to a linear polarized thermal radiation, we calculate the circular polarization intensity, and discuss its possible relevance to the circular polarization intensity of  the Cosmic Microwave Background radiation.

\end{abstract}

PACS: 73.50.Fq, 42.50.Xa, 98.70.Vc

\vskip0.1cm
\noindent
{\bf Introduction.}
\hskip0.1cm
Modern cosmological observations of the cosmic microwave
background (CMB) radiation provide important
evidences to understand our Universe. 
Cosmological informations encoded in the CMB radiation concerns not only
temperature fluctuations and the spectrum of anisotropy pattern, but
also the intensity and spectrum of linear and circular
polarizations. It is generally expected that some relevant linear
and circular polarizations of CMB radiation should be present,
and polarization fluctuations are smaller than temperature
fluctuations  \cite{nature}. Recently, there are several ongoing experiments \cite{exp1} to attempt to
measure CMB polarizations. Theoretical studies of CMB
polarizations were carried out in Refs.~\cite{kaiser1983,cosowsky1994}, and
numerical calculations \cite{num1,num2} have confirmed that about
$10\%$ of the CMB radiation fields are linear polarizations, via the
Compton and Thompson scatterings of unpolarized photons at the
last scattering surface (the redshift $z\sim 10^3$). 
It is important
from theoretical points of view to understand the generation of
CMB linear and circular polarizations.
\comment{
In order to describe polarizations of radiation fields, one
uses the Stokes parameters $I, Q, U,$ and $V$, where $I$ describes
the intensity of radiation field, $Q$ and $U$ linear polarization
intensities and $V$ circular polarization intensity of radiation
fields \cite{jackson}.
}

In principle, under effects of background fields, particle
scatterings and temperature fluctuations, linear polarizations
of CMB radiation field propagating from the last scattering
surface can rotate each other and convert to circular
polarizations. This is described by the formalism of Faraday
rotation (FR) and conversion (FC) \cite{faraday con}, and the
conversion from linear to circular polarizations is given by
the time evolution of the Stokes parameter $V$:
\bea
\dot{V}=2U\frac{d}{dt}(\Delta\phi_{FC})
\label{fc},
\eea 
where $\Delta\phi_{FC}$ is the Faraday conversion phase shift \cite{fc}.

Refs.~\cite{fc,khodam} present the role of background magnetic fields in
producing the CMB circular polarization,
and $\Delta\phi_{FC} \sim 10^{-19}$ for micro gauss
magnetic fields. In refs.~\cite{gio1,gio2},
the angular power spectrum of CMB
circular polarizations and relevant correlations are studied
in the case that circular polarization is generated
by photon-electron scattering in the presence of magnetic fields.
I is shown that Lorentz symmetry
violation in some extension of standard model for particle physics \cite{khodam,lv cos} and an axion-like
cosmological pseudoscalar field \cite{axion} can generate the CMB circular polarization. 
Ref.~\cite{khodam} shows that the noncommutative QED with the
Seiberg-Witten expansion of fields in the last scattering surface can also generate the CMB circular polarization, and the FC phase shift $\Delta\phi_{FC}\sim 10^{-17} $. 

In this letter we show that circular polarizations of radiation
fields can be generated from the effective Euler-Heisenberg
Lagrangian 
(see review articles in refs.~\cite{WH2000,dunne,xue}). By
taking into account this effective Lagrangian, and using the
Quantum Boltzmann Equation \cite{cosowsky1994}, we study the
time-evolution of the Stokes parameter $V$. Applying our results to the
homogeneous CMB radiation with non-vanishing linear polarizations,
we obtain the upper limit of circular polarization intensity $V/T_0 < 7\times
10^{-10}$ in units of thermal temperature $T_0$ of the CMB radiation,
and the corresponding $\Delta\phi_{FC}< 1.6\times 10^{-13}$. 


\vskip0.1cm
\noindent
{\bf Stokes parameters.}
\hskip0.1cm
A nearly monochromatic electromagnetic wave propagating in
the $\hat z$-direction 
is described by:
 \bea
E_x=a_x(t)\cos[\omega_0t-\theta_x(t)],\quad
  E_y=a_y(t)\cos[\omega_0t-\theta_y(t)],\eea
where amplitudes $a_{x,y}$ and
phase angles $\theta_{x,y}$ are slowly varying functions with respect to the period ${\mathcal T}_0=2\pi/\omega_0$.
Any correlation between the $a_x$- and $a_y$-components indicates polarizations of electromagnetic waves.
In a classical description \cite{jackson}, Stokes parameters, which describe polarization states of a nearly monochromatic electromagnetic wave, 
are defined as the following time averages: \bea I_c &=&\langle a^2_
x \rangle+ \langle a^2_y\rangle, \nonumber\\ Q_c &=&\langle a^2_ x
\rangle- \langle a^2_y\rangle, \nonumber\\ U_c &=&\langle 2a_ x
a_y\cos(\theta_x-\theta_y)\rangle,\nonumber\\ V_c &=&\langle 2a_ x
a_y\sin(\theta_x-\theta_y)\rangle,
\label{ic}
\eea
where the parameter $I_c$
is total intensity, $Q_c$ and $U_c$ intensities of linear
polarizations of electromagnetic  waves, whereas the $V_c$
parameter indicates
the difference between left- and right- circular polarizations
intensities. Linear polarization can also be characterized through
a vector of modulus $P_L\equiv\sqrt{Q_c^2+U_c^2}$.
It is important to notice that the Stocks parameters (\ref{ic}) are defined for a monochromatic electromagnetic wave with a definite momentum $k$.
Given a linear polarization,
one can always transform them to a coordinate system where $Q_c$ or $U_c$
vanishes leaving no circular polarization $V_c=0$. In order to generate a net circular
polarization via birefringence there must be some special coordinate
system so that one linear polarization state propagates differently from the other due to interactions.

In a quantum-mechanical description, Stokes parameters can be
equivalently defined as follows. An arbitrary polarized
state of a photon $(|k_0|^2=|{\bf k}|^2)$, propagating in the
$\hat z$-direction, is given by \bea
|\epsilon\rangle=a_1\exp(i\theta_1)|\epsilon_1\rangle+a_2\exp(i\theta_2)|\epsilon_2\rangle,\eea
where linear bases $|\epsilon_1\rangle$ and
$|\epsilon_2\rangle$ indicate the polarization states in the $x$-
and $y$-directions. Quantum-mechanical operators in this
linear bases, corresponding to Stokes parameter, are given by
\bea
\hat{I}&=&|\epsilon_1\rangle\langle\epsilon_1|+|\epsilon_2\rangle\langle\epsilon_2|,\nonumber\\
\hat{Q}&=&|\epsilon_1\rangle\langle\epsilon_1|-|\epsilon_2\rangle\langle\epsilon_2|,\nonumber\\
\hat{U}&=&|\epsilon_1\rangle\langle\epsilon_2|+|\epsilon_2\rangle\langle\epsilon_1|,\nonumber\\
\hat{V}&=&i|\epsilon_2\rangle\langle\epsilon_1|-i|\epsilon_1\rangle\langle\epsilon_2|.
\label{i-v} \eea An ensemble of photons in a general mixed state
is described by a normalized density matrix $\rho_{ij}\equiv
(\,|\epsilon_i\rangle\langle \epsilon_j|/{\rm tr}\rho)$, and the dimensionless
expectation values for Stokes parameters are given by
\bea
I\equiv\langle  \hat I \rangle &=& {\rm tr}\rho\hat I
=1,\label{i}\\
Q\equiv\langle  \hat Q \rangle &=& {\rm tr}\rho\hat{Q}=\rho_{11}-\rho_{22},\label{q}\\
U\equiv\langle
 \hat U\rangle &=&{\rm tr}\rho\hat{U}=\rho_{12}+\rho_{21},\label{u}\\
V\equiv\langle  \hat V \rangle &=& {\rm
tr}\rho\hat{V}=i\rho_{21}-i\rho_{21}, \label{v}
\eea
where ``$\rm tr$'' indicates the trace in the space of polarization states. This shows
the relationship between four Stokes parameters and the
$2\times 2$ density matrix $\rho$ for photon polarization states.



\vskip0.1cm
\noindent
{\bf Euler-Heisenberg Lagrangian and circular polarizations.}
\hskip0.1cm
The Euler-Hesinberg effective Lagrangian 
is given as follows: 
\bea \pounds_{eff} = \pounds_0 +
\delta\pounds,
\label{maxwell} 
\eea 
where the first term $\pounds_0=
-\frac{1}{4}F_{\mu\nu}F^{\mu\nu}$ is the
classical Maxwell Lagrangian, 
and the second term $\delta\pounds$
\bea
\delta \pounds &\approx &
\frac{\alpha^2}{90m^4}
\left[(F_{\mu\nu}F^{\mu\nu})^2
+\frac{7}{4}(F_{\mu\nu}\tilde{F}^{\mu\nu}
)^2\right],
\label{eh}
\eea
where $m$ is the electron mass,
$\tilde{F}^{\mu\nu}=\epsilon^{\mu\nu\alpha\beta}F_{\alpha\beta}$ (see
review articles in refs.~\cite{WH2000,dunne,xue}). 

We express the electromagnetic field strength $F_{\mu\nu}=\partial_\mu A_\nu-\partial_\nu A_\mu$, and 
free gauge field $A_\mu$ in terms of plane wave solutions in the Coulomb gauge \cite{zuber}, 
 \beq A_\mu(x) = \int \frac{d^3 k}{(2\pi)^3
2 k^0} \left[ a_i(k) \epsilon _{i\mu}(k)
        e^{-ik\cdot x}+ a_i^\dagger (k) \epsilon^* _{i\mu}(k)e^{ik\cdot x}
        \right],\eeq
where $\epsilon _{i\mu}(k)$ are the polarization
four-vectors and the index $i=1,2$, representing two transverse polarizations of a free photon with four-momentum $k$ and $k^0=|{\bf{k}}|$. $a_i(k)$ 
$[a_i^\dagger (k)]$ are the creation [annihilation] operators, which satisfy the canonical commutation relation
\begin{equation}
        \left[  a_i (k), a_j^\dagger (k')\right] = (2\pi )^3 2k^0\delta_{ij}\delta^{(3)}({\bf k} - {\bf k}' ).
\label{comm}
\end{equation}
The density operator describing an ensemble of free photons in the space of energy-momentum and polarization state is given by 
\bea
\hat\rho=\frac{1}{\rm {tr}(\hat \rho)}\int\frac{d^3p}{(2\pi)^3}
\rho_{ij}(p)a^\dagger_i(p)a_j(p),
\eea
where $\rho_{ij}(p)$ is the general density-matrix (\ref{i}-\ref{v}) in the space of polarization states with a fixed energy-momentum ``$p$''.
The number operator $
D^0_{ij}(k)\equiv a_i^\dag (k)a_j(k)$ and its
expectation value is defined by
\bea
\langle\, D^0_{ij}(k)\,\rangle\equiv {\rm tr}[\hat\rho
D^0_{ij}(k)]=(2\pi)^3 \delta^3(0)(2k^0)\rho_{ij}(k).
\eea
The time evolution of photon polarization states is related to the time evolution of the density matrix $\rho_{ij}(k)$, which is governed by the following Quantum Boltzmann Equation (QBE) \cite{cosowsky1994},
\bea
(2\pi)^3 \delta^3(0)(2k^0)
\frac{d}{dt}\rho_{ij}(k) = i\langle \left[H^0_I
(t);D^0_{ij}(k)\right]\rangle-\frac{1}{2}\int dt\langle
\left[H^0_I(t);\left[H^0_I
(0);D^0_{ij}(k)\right]\right]\label{bo}\rangle,
\eea
where the interacting Hamiltonian $H^0_I(t)=-\delta\pounds$ (\ref{eh}).  The first term on the right-handed side is a
forward scattering term, and the second one is a higher order collision term.

It is known that the linear Maxwell Lagrangian $\pounds_0$ in Eq.~(\ref{maxwell}) does not generate circular polarizations. We attempt to compute the effect of the non-linear Euler Heisenberg Lagrangian (\ref{eh}) on the
generation of circular polarizations by using QBE (\ref{bo}). 
\comment{
The effective Euler-Heisenberg Lagrangian
(\ref{eh}) corresponds to the interacting Hamiltonian, 
\bea
H^0_I(t) &=& -\frac{\alpha^2}{90m^4}\int d^3x
\left[(F_{\mu\nu}{F}^{\mu\nu} )^2+\frac{7}{4}(F_{\mu\nu}\tilde{F}^{\mu\nu} )^2\right],
\label{eh}
\eea
}
Eq.~(\ref{eh}) is perturbatively small, at the order of $\alpha^2$, so that we only
compute the first
order of QBE, i.e.~the first term in r.h.s.~of Eq.~(\ref{bo}), and neglect the second term which is of the order of $\alpha^4$. The contribution from the first term $(F_{\mu\nu}{F}^{\mu\nu} )^2$ in Eq.~(\ref{eh}) vanishes, because it is a squared Maxwell action and commutates with the number operator $D^0_{ij}$. While the second nonlinear term $(F_{\mu\nu}\tilde{F}^{\mu\nu} )^2$ does not commutate with the number operator $D^0_{ij}$ and gives non-vanishing contributions.

As a result, we approximately obtain the time-evolution equation
for the density matrix, \bea (2\pi)^3 \delta^3(0)2k^0
\frac{d}{dt}\rho_{ij}(k) \!\!&\approx& i\langle
\left[H^0_I(t),D^0_{ij}(k)\right]\rangle\nonumber\\
&=&\frac{56\alpha^2}{45m^4}(2\pi)^3\delta^3(0)\epsilon^{\mu\nu\alpha\beta}\epsilon^{\sigma\nu'\gamma\beta'}k_\gamma
k_\mu[\epsilon_{s\nu}(k)\epsilon_{l'\beta'}(k)]\nonumber\\
&\times&\left[\rho_{l'j}(k)\delta^{si}-\rho_{is}(k)\delta^{l'j}+\rho_{sj}(k)\delta^{l'i}-\rho_{il'}(k)\delta^{sj}\right]\nonumber\\
&\times&\!\!\! \int\frac{d^3p}{(2\pi)^32p^0}p_\alpha
p_\sigma[\epsilon_{s'\beta}(p)\epsilon_{l\nu'}(p)][\rho_{ls'}(p)+\rho_{s'l}(p)+\delta^{s'l}].
\label{j1}
\eea
The calculations are tedious, but straightforward. We first apply
the Wick theorem to arrange all creation operators to the left and
all annihilation operators to the right, then we use the contraction
rule \bea \langle \, a^\dag_{s'}(p')a_{s}(p)\, \rangle
&=&2p^0(2\pi)^3\delta^3(\mathbf{p}-\mathbf{p'})\rho_{ss'}(p),
\label{contraction} \eea to calculate all possible contractions of
creation and annihilation operators $a_i^\dag$ and $a_j$. For
example, we calculate the
expectation value 
\bea
\langle p|\, a^\dag_{s'}(p')a_{s}(p)a^\dag_{l'}(q')a_{l}(q)\,|p \rangle
\!\!\!&=& \!\!\!\langle p|\,
a^\dag_{s'}(p')a^\dag_{l'}(q')a_{s}(p)a_{l}(q)\,|p \rangle\nonumber\\
&+&\!\!\!2p^0(2\pi)^3\delta^{sl'}\delta^3(\mathbf{p}-\mathbf{q'})\langle p|\, a^\dag_{s'}(p')a_{l}(q)\,|p \rangle\nonumber\\
&=&\!\!\!\ 4p^0q^0(2\pi)^6\delta^3(\mathbf{p}-\mathbf{p'})
\delta^3(\mathbf{q}-\mathbf{q'})\rho_{ss'}(p)\rho_{ll'}(q)\nonumber\\
&+&\!\!\!
4p^0q^0(2\pi)^6\delta^3(\mathbf{p}-\mathbf{q'})
\delta^3(\mathbf{q}-\mathbf{p'})\rho_{s'l}(q)[\delta_{sl'}+\rho_{sl'}(p)],\nonumber
\eea
where the first line results from the Wick theorem and commutation relations (\ref{comm}), while the second line results from all possible contractions (\ref{contraction}) of operators $a_s^\dag$ and $a_s$.
Using Eq.~(\ref{j1}), we obtain the time-evolutions for  Stocks parameters (\ref{i}-\ref{v}) as follows:
\bea
\dot{I}(k)
&=&0,\label{id}\\
\dot{Q}(k)
&=&\hat X\Big\{[\rho_{21}(k )-\rho_{12}(k )][\rho_{21}(p)-\rho_{12}(p)]\nonumber\\
&\times&
[\epsilon_{2\nu}(k)\epsilon_{1\beta'}(k)\epsilon_{1\beta}(p)\epsilon_{2\nu'}(p)
+\epsilon_{1\nu}(k)\epsilon_{2\beta'}(k)\epsilon_{2\beta}(p)\epsilon_{1\nu'}(p)]\Big\}\nonumber\\
&=&\!\!-\hat X\Big\{V(k )V(p)
[\epsilon_{2\nu}(k)\epsilon_{1\beta'}(k)\epsilon_{1\beta}(p)\epsilon_{2\nu'}(p)
+\epsilon_{1\nu}(k)\epsilon_{2\beta'}(k)\epsilon_{2\beta}(p)\epsilon_{1\nu'}(p)]\Big\},\label{qd}\\
\dot{U}(k)
&=&\hat
X\Big\{[\rho_{21}(k )-\rho_{12}(k )][\rho_{11}(p)-\rho_{22}(p)]
\epsilon_{2\nu}(k)\epsilon_{2\beta'}(k)\epsilon_{1\beta}(p)\epsilon_{1\nu'}(p)\Big\}\nonumber\\&=&
\hat X\Big\{iV(k )Q(p)
\epsilon_{2\nu}(k)\epsilon_{2\beta'}(k)\epsilon_{1\beta}(p)\epsilon_{1\nu'}(p)\Big\},
\label{ud}\\
\dot{V}(k)
&=&\hat X\Big\{[\rho_{22}(k )-\rho_{11}(k )][\rho_{12}(p)+\rho_{21}(p)]\epsilon_{1\nu}(k)
\epsilon_{1\beta'}(k)\epsilon_{2\beta}(p)\epsilon_{2\nu'}(p)\nonumber\\
&+&[\rho_{21}(k )-\rho_{12}(k )]
[\rho_{11}(p)-\rho_{22}(p)]
\epsilon_{2\nu}(k)\epsilon_{2\beta'}(k)\epsilon_{1\beta}(p)\epsilon_{1\nu'}(p)\Big\}\nonumber\\
&=&\hat X\Big\{Q(k )U(p)\epsilon_{1\nu}(k)
\epsilon_{2\beta'}(k)\epsilon_{2\beta}(p)\epsilon_{1\nu'}(p)\nonumber\\
&+&iV(k )Q(p)
\epsilon_{2\nu}(k)\epsilon_{2\beta'}(k)\epsilon_{1\beta}(p)\epsilon_{1\nu'}(p)\Big\},
\label{vd} \eea
where $k$ indicates the energy-momentum state of incoming photons in a radiation field,  and $p$ indicates the energy-momentum states of virtual photons in vacuum,
and  the operator $\hat X$ is defined as following integral
overall energy-momentum states $p$, \bea \hat
X\Big\{\cdot\cdot\cdot\Big\}\equiv\frac{16\times7\alpha^2}{45m^4k^0}\int\frac{d^3p}{(2\pi)^32p^0}\left
[\epsilon^{\mu\nu\alpha\beta}\epsilon^{\sigma\nu'\gamma\beta'}k_\gamma
k_\mu p_\alpha p_\sigma \right]\Big\{\cdot\cdot\cdot\Big\}.
\label{xint}\eea
The $I$ modes represent the ensemble of photons.
The $Q$ and $U$ modes represent the ensemble of linearly polarized photons, and the $V$ mode represents the ensemble of circularly polarized photons. 

Eq.~(\ref{eh}) gives an interacting vortex of four photons. Eqs.~(\ref{id}-\ref{vd}) result from the tadpole diagram of a photon loop integrating all contributions ``$p$'' of virtual photons in vacuum (see Eq.~\ref{xint}). This indicates that polarization states of a propagating photon with momentum ``$k$'' interact with those of virtual photons in vacuum. If the photon ``$k$'' is not initially polarized, i.e., $Q(k)=U(k)=V(k)=0$, then Eqs.~(\ref{id}-\ref{vd}) show that the photon ``$k$'' propagating through vacuum does not acquire polarizations. Instead, the photon ``$k$'' is linearly polarized,  as if there were a particular orientation of local magnetic field. Interacting with this local magnetic field, virtual photons can develop circular polarization states, in turn these states back-react with polarization states of the photon ``$k$''.  As a result, Eqs.~(\ref{id}-\ref{vd}) show that the photon ``$k$'' acquires a net circular polarization. 

It is important to notice that in the right-handed side of Eq.~(\ref{vd}), the linearly polarized modes $Q(k)$ and $U(p)$ are in different momentum states ``$k$'' and ``$p$'' so that they are independent modes. This cannot be made by a coordinate transformation so that one of them ($U$ or $V$) vanishes, as discussed for the Stocks parameters (\ref{ic}) or (\ref{i}-\ref{v}) for a monochromatic electromagnetic wave with a definite momentum ``$k$''.
In addition, the right-handed sides of Eqs.~(\ref{qd},\ref{ud},\ref{vd}) show nonlinear interactions between $Q$, $U$ and $V$ modes of a given momentum state ``$k$'' of a radiation field and all possible momentum states ``$p$'' from vacuum contributions. In these nonlinear interactions, $Q$, $U$ and $V$ modes differently interact with each other
leading to circular polarizations, i.e., non-vanishing $V(k)$ modes, provided $Q(k)$ and $U(p)$ are non zero.

In Eq.~(\ref{id}), $\dot I=0$ implies in the ensemble of photons, the total intensity of photons is constant in time-evolution. In Eqs.~(\ref{qd},\ref{ud}) and (\ref{vd}), the time-evolution $\dot Q$, $\dot U$ and $\dot V$ are given by the combinations of $Q$, $U$ and $V$ modes, which indicates a rotation or conversion between these modes as long as the effective interaction (\ref{eh}) acts.  The time-evolution $\dot V$ is proportional to $Q$ and $U$ modes. This indicates that an ensemble of linearly polarized photons will acquire
circular polarizations due to
the Euler-Heisenberg Lagrangian (\ref{eh}).
We are interested in considering an initially linear polarized electromagnetic radiation propagating through vacuum, and calculating how much the intensity of circular polarization can be converted from the intensity of linear polarization, due to the non-linear Euler-Heisenberg interaction.


\vskip0.1cm
\noindent
{\bf Intensity of circular polarizations in CMB.}
\hskip0.1cm
We apply our results (\ref{id}-\ref{vd}) to the ensemble of thermal CMB
photons, $f_{\rm BB}(p) = 1/(e^{p/T}-1)$, where the temperature $T$ and photon momentum $p$ are in the comoving frame.
Thus photon energy and number densities are given by: 
\bea
\varepsilon_\gamma = 2\int \frac{d^3p}{(2\pi)^3}\, p\,f_{\rm
BB}(p)= \frac{\pi^2}{15}\,T^4, \quad n_\gamma = 2\int
\frac{d^3p}{(2\pi)^3}f_{\rm BB}(p)= \frac{2\zeta(3)}{\pi^2}\,T^3.
\label{nd} 
\eea 
and the mean energy for each thermal photon
\begin{eqnarray}
\frac{\varepsilon_\gamma}{n_\gamma}&=&\frac{\pi^4\,T}{[30\zeta(3)]}
\approx 2.7\, T,
\label{neps}
\end{eqnarray}
corresponds to the intensity $I_c$ (\ref{ic}). We consider that
the thermal radiation is initially polarized and propagates
through vacuum. In order to calculate the final intensity of
circular polarizations (\ref{vd}), we approximate \bea Q(k)
\epsilon_{2\beta'}(k)\epsilon_{1\nu}(k)\approx C_Qf_{\rm
BB}(k)\delta_{\beta\nu'},\quad U(p)
\epsilon_{2\beta}(p)\epsilon_{1\nu'}(p)\approx C_U f_{\rm
BB}(p)\delta_{\beta\nu'}, \label{aq} 
\eea 
where coefficient $C_Q
(C_U)$ is the ratio of linear $Q (U)$-polarization intensity and
total intensity. The coefficient $C_Q$ represents the linear polarization of the real photon ``$k$'' propagating in vacuum $C_Q < 1$, while $C_U$ represents the sum over all contributions of linear polarization of virtual photons ``$p$'' in vacuum, and $C_U\simeq 1$.  $C_Q$ and $C_U$ are
independent of each other, because they are associated to different
momentum states $k$ and $p$ of photons. 

Assuming that the converted intensity of circular polarization  is
much smaller than the intensity of linear polarization, we neglect
the second term in the right-handed side of Eq.~(\ref{vd}).
Integrating Eq.~(\ref{vd}) over all momentum states $p$ [see
Eq.~(\ref{xint})], we approximately obtain \bea \dot{V}(k)\cong
\Big(\frac{\pi^2\alpha^2}{2}\Big)\Big(\frac{T}{m}\Big)^4
C_UC_Q[kf_{\rm BB}(k)], \label{vda} \eea where $k$ is the momentum
of thermal photons. The form of  expression (\ref{vda}) can be
understood as the rate of converting the linearly polarized mode to
a circularly polarized mode. The linearly polarized $Q$-mode of
momentum state $k$ interacts with linearly polarized $U$-modes of
all momentum states $p$ in vacuum, and converts to the circularly
polarized $V$-mode of momentum state $k$.  
The factor $[kf_{\rm BB}(k)]$ is due to
the energy-spectrum of thermal photons and $(T/m)^4$ comes from
the summation over all momentum states $p$ (\ref{nd}). Integrating
Eq.~(\ref{vda}) overall energy-momentum states ``$k$'' of the
black-body distribution $f_{\rm BB}(k)$, and normalizing it by the
total number-density $n_\gamma$ of thermal photons, we obtain \bea
\frac{d V}{d t}\cong
\Big(\frac{\pi^2\alpha^2}{2}\Big)\Big(\frac{T}{m}\Big)^4\Big(\frac{\varepsilon_\gamma}{n_\gamma}\Big)C_UC_Q
,\quad V\equiv\frac{2}{n_\gamma}\int \frac{d^3k}{(2\pi)^3}V(k).
\label{vdaf} 
\eea 
Finally, multiplying the mean intensity
$\varepsilon_\gamma/n_\gamma $, we obtain \bea \frac{d
V}{dt}&\cong &
\Big(\frac{\pi^2\alpha^2}{2}\Big)\Big(\frac{T}{m}\Big)^4
\Big(\frac{\varepsilon_\gamma}{n_\gamma}\Big)^2C_UC_Q;
\label{result1} 
\eea 
analogously, we redefine the Stokes parameters $Q$ and
$U$ in energy units by multiplying the mean energy
$\varepsilon_\gamma/n_\gamma$, 
\begin{eqnarray} 
 U=\Big(\frac{\varepsilon_\gamma}{n_\gamma}\Big)
\frac{2}{n_\gamma}\int \frac{d^3p}{(2\pi)^3}U(p)\approx
C_U\Big(\frac{\varepsilon_\gamma}{n_\gamma}\Big),
\label{cqu}
\end{eqnarray}
and $U\Rightarrow Q$, corresponding to those in Eq.~(\ref{i-v}).

To estimate the $V$, we integrate over the comoving time $\int dt=\int dz/H(z)$, where the redshift $z\in [0, 10^3]$, the Hubble function $H(z)=H_0[\Omega_M(z+1)^3+\Omega_\Lambda)]^{1/2}$ for $\Omega_M\simeq 0.3$, $\Omega_\Lambda\simeq 0.7$ and $H_0=75$ km/s/Mpc, and the temperature 
$T = T_0(1+z)$ [$T_0\approx 2.725 K^\circ=2.349\times 10^{-4}{\rm
eV}=(0.511{\rm cm})^{-1}$] in the standard cosmology \cite{eric,bbody}. From Eqs.~(\ref{neps},\ref{result1}) we obtain in units of the present CMB temperature $T_0$, 
\begin{eqnarray}
\frac{\Delta
V}{T_0}&\simeq &7.3
\Big(\frac{\pi^2\alpha^2}{2}\Big)\Big(\frac{T_0}{m}\Big)^4
T_0C_U\,C_Q\int^{1000}_{0}dz(1+z)^{6}/H(z)\nonumber\\
&\approx & 7\times 10^{-10} C_U\,C_Q < 7\times 10^{-10}. \label{result0}
\end{eqnarray}
\comment{
As a preliminary result, we apply Eq.~(\ref{result1}) to the
homogeneous CMB radiation at a temperature $T\approx
2.725K^\circ=2.349\times 10^{-4}{\rm eV}=(0.511{\rm cm})^{-1}$
\cite{bbody}. Considering time-variation $\Delta t\sim  L$ and
$L\sim 10 {\rm Gpc}=3.1\times 10^{28}$cm for red-shift $z\sim
1000$, we obtain $(T/m)^4 =4.5\times 10^{-38}$, $TL\approx
6.1\times 10^{28}$ and}
We find that this intensity of circular polarizations is very
small, as compared with the CMB anisotropy $\Delta T/T_0\approx
10^{-5}$ \cite{anisotropy}, which measures the inhomogeneity of the
CMB radiation. 
From Eqs.~(\ref{fc}), (\ref{result1}) and (\ref{cqu})
, and we obtain the Faraday conversion phase shift
\bea 
\Delta\phi_{FC} &\simeq &1.35
\Big(\frac{\pi^2\alpha^2}{2}\Big)\Big(\frac{T_0}{m}\Big)^4
T_0 C_Q\int^{1000}_{0}dz(1+z)^{5}/H(z)\nonumber\\
&\approx & 1.6\times 10^{-13} C_Q < 1.6\times 10^{-13}.
\label{result2}
\eea
Our results (\ref{result0},\ref{result2}) give the upper limit of the
effects of  non-linear Euler-Heisenberg Lagrangian on the intensity of
CMB circular polarization. 

Analogously to Ref.~\cite{saw}, which also discusses the effect of photon-photon scatterings on the CMB circular polarization, we take the linear polarization coefficient $C_Q\approx 3 \times 10^{-5}$ at the last scattering surface, corresponding to $\Delta T_{Q}/T_0$ with the maximum $\Delta T_{Q}\approx 3\times 10^{-5} K^\circ$ in the WMAP data \cite{WMAP}, we obtain $\Delta V /T_0 \approx 2.1\times 10^{-14}$, and $\Delta\phi_{FC}\approx 4.8\times 10^{-18}$. 
This result implies that the Euler-Heisenberg effect on the CMB circular polarization could be rather important, compared with those effects mentioned in the introductory paragraph.


\vskip0.1cm
\noindent
{\bf Conclusion and remarks.}
\hskip0.1cm
In this letter, by approximately solving the first order of Quantum
Boltzmann Equation for the density matrix of a photon ensemble, and
time-evolution of Stokes parameters, we show that
propagating photons convert their linear polarizations to circular
polarizations by the nonlinear Euler-Heisenberg interactions. 
We discussed this Euler-Heisenberg effect on the circular polarization of CMB photons, and showed that this effect is very small, as compared with the present CMB temperature $T_0$. Nevertheless, observational studies on such circular polarization are clearly warranted.
What and how do
we need to observe? Is non-zero circular polarization at one point
in the sky enough,
or should there be correlations with the pattern of linear polarization? To our knowledge, there is not any current
polarization experiment that directly measures the CMB circular polarization. However,
in the next five years considerably more detailed information about the CMB polarization will be
delivered by the Planck satellite \cite{plank} and ground based,
high resolution polarization experiments such as $\rm{ACTPol}$
\cite{act}, $\rm{PIXIE}$ \cite{pix}, $\rm{SPIDER}$ \cite{spider},
$\rm{PolarBear}$ \cite{polar}, and $\rm{SPTPol}$ \cite{sp}. 
The sensitivity (polarization) ${\Delta T}_Q/{T_0}$ of the Planck satellite for hight and low frequency is in the order of $10^{-6}$.
This seems to be still far from the CMB circular polarization 
$\Delta V /T_0 \approx 2.1\times 10^{-14}$ estimated in this letter.
On the other hand, it would be interesting to see this Euler-Heisenberg effect on the circular polarization of laser photons \cite{imanxue_future}.
\comment{
To end this article, we make a remark on this circular polarization effect in an $X$-ray laser experiment.
Our results (\ref{id}-\ref{vd}) imply that a linear polarized laser beam must develop circular polarizations while it is
propagating through vacuum. In this case, the amplitudes of linear and circular polarizations are comparable, one needs to completely solve Eqs.~(\ref{id}-\ref{vd}) and respectively obtains the intensities of linear and circular polarizations of the laser beam \cite{imanxue_future}.
}

\vskip0.1cm
\noindent
{\bf Acknowledgment.}
\hskip0.1cm
One of the authors, I.~Motie thanks Professor R.~Ruffini for his hospitality, during his stay at ICRANet Pescara, Italy, where this work is done. We thank the anonymous referee for his comments and suggestions. 


\end{document}